\documentclass{article}
\usepackage[utf8]{inputenc}
\usepackage[colorlinks,urlcolor=blue,linkcolor=blue,citecolor=blue]{hyperref}
\usepackage{color,array}
\usepackage[english]{babel} 
\usepackage{subfig}

\usepackage{amssymb}
\usepackage{amsmath}
\usepackage{txfonts}
\usepackage{mathdots}
\usepackage[classicReIm]{kpfonts}
\usepackage{graphicx}
\usepackage{algorithm}
\usepackage{float}
\usepackage[square,sort,comma,numbers]{natbib}

\usepackage[noend]{algpseudocode}
\usepackage{indentfirst}
\usepackage[preprint]{neurips_2020}

\usepackage[font=small, labelsep = period]{caption}
\title{Real-time Interface Control with Motion Gesture Recognition based on Non-contact Capacitive Sensing}
\author{%
  Hunmin Lee \\
  Department of Computer Science\\
  Georgia State University\\
  Atlanta, GA 30302 \\
  \texttt{hlee185@student.gsu.edu} \\
   \And
   Jaya Krishna Mandivarapu \\
   Department of Computer Science\\
   Georgia State University\\
   Atlanta, GA 30302 \\
   \texttt{jmandivarapu1@student.gsu.edu} \\
   \And
   Nahom Ogbazghi\\
   Department of Computer Science\\
   Georgia State University\\
   Atlanta, GA 30302 \\
   \texttt{nogbazghi1@student.gsu.edu} \\
   \And
    Yingshu Li \\
   Department of Computer Science\\
   Georgia State University\\
   Atlanta, GA 30302 \\
   \texttt{yili@gsu.edu} \\
}
\date{November 2021}

\begin{document}

\maketitle
\begin{abstract} 
Capacitive sensing is a prominent technology that is cost-effective and low power consuming with fast recognition speed compared to existing sensing systems. On account of these advantages, Capacitive sensing has been widely studied and commercialized in the domains of touch sensing, localization, existence detection, and contact sensing interface application such as human-computer interaction. However, as a non-contact proximity sensing scheme is easily affected by the disturbance of peripheral objects or surroundings, it requires considerable sensitive data processing than contact sensing, limiting the use of its further utilization. In this paper, we propose a real-time interface control framework based on non-contact hand motion gesture recognition through processing the raw signals, detecting the electric field disturbance triggered by the hand gesture movements near the capacitive sensor using adaptive threshold, and extracting the significant signal frame, covering the authentic signal intervals with 98.8\% detection rate and 98.4\% frame correction rate. Through the GRU model trained with the extracted signal frame, we classify the 10 hand motion gesture types with 98.79\% accuracy. The framework transmits the classification result and maneuvers the interface of the foreground process depending on the input. This study suggests the feasibility of intuitive interface technology, which accommodates the flexible interaction between human to machine similar to Natural User Interface, and uplifts the possibility of commercialization based on measuring the electric field disturbance through non-contact proximity sensing which is state-of-the-art sensing technology.

\end{abstract}

\section{Introduction} 

\begin{figure}
\centering{\includegraphics[width=\textwidth]{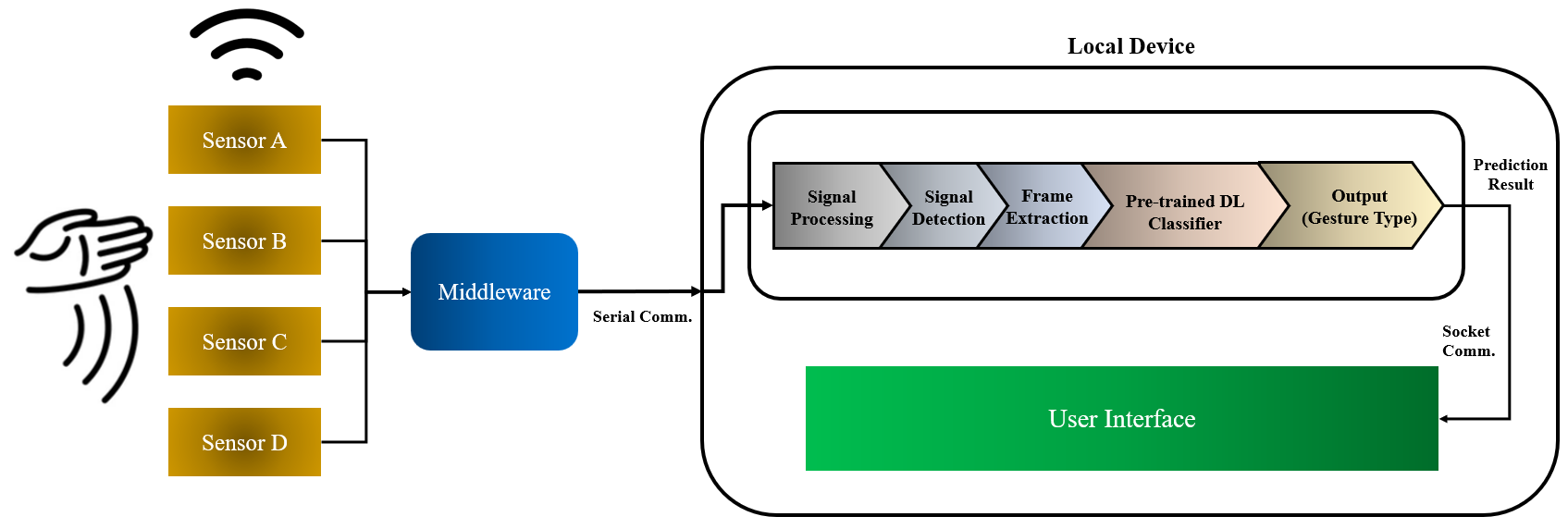}}
\caption{Framework overview: Our proposed Motion Gesture Recognition System using non-contact capacitive sensing.}
\label{SystemOverview}
\end{figure}

Capacitive proximity sensing quantitatively measures the variation of Electric Field (EF) around a sensor based on the EF disturbance which is physically triggered from a dialectic object or conductor nearby. Through EF disturbance caused by peripheral objects, it provides a novel methodology to be utilized in diverse domains by detecting target signals and controlling the machines through digitalization. Among the two main capacitive sensing mechanisms [1, 2]: contact sensing and non-contact sensing, our focus lies on non-contact proximity sensing. In this paper, we propose an automated real-time hand motion gesture detection and classification system which controls the application’s User Interface (UI) where the gesture classification output results will be used to control the UI. The architecture is composed of the following activities: extracting raw signals from sensors, signal processing, motion gesture detection, gesture frame extraction, gesture classification with a GRU classifier, and socket communication to control the foreground process of UI. Each step in the framework is selected with an optimal mechanism after thorough step-wise analysis in order to acquire efficiency and feasibility, considering the future utilization in applications to control their UI through physical hand motion gestures. In our experiments, we collected the hand motion gesture data from the capacitive sensors and derived 10 gesture type datasets (totally 1000 datasets) from three subjects in our GitHub [3] to be accessed by the public. Our experimental results validate our approach with a 98.8\% detection rate, 98.4\% extraction rate, and 98.79\% accuracy in classification. Fig.\ref{SystemOverview} displays the overall pipeline of our proposed system.\\

The main contributions of this paper are as follows:\\
1. This paper presents an end-to-end pipeline and step-wise evaluation for an automated \& real-time interface control system with hand motion gesture recognition based on non-contact capacitive sensing. To evaluate the performance of the proposed architecture, multiple experiments were conducted over the datasets acquired for this scenario on three metrics: detection rate, frame extraction rate, and gesture classification accuracy.\\
2. Our end-to-end system presents the controllability of User Interface which suggests the feasibility of new product commercialization in the domains of IoT, HCI (Human-Computer Interaction), NUI (Natural User Interface), and HMI (Human Machine Interface) through non-contact proximity sensing with low cost, low computation, and low latency compared to the existing sensing mechanisms.\\
3. We open-sourced our established benchmark dataset and algorithm design through Github [3] for further development progress of capacitive sensing systems through crowdsourcing.\\

This paper is organized as follows. In further parts of this section, we address the footprints of non-contact capacitive sensing applications by reviewing the existing studies. In Section 2, we introduce our methodology, implementing signal processing mechanisms by thoroughly analyzing the extracted signals from our designed capacitive sensing system from multiple perspectives. We also propose the adaptive threshold in order to detect and extract the signal frame which covers the relevant signals. Section 3 presents the Deep Learning classifier to train and categorize the signals into the right gesture. Furthermore, we implement the classifier to control the UI and suggest its potential usability. In section 4, we evaluate our system ramification performance through conducting multiple experiments with three metrics. Finally, we conclude our paper by presenting conclusions and future works.

\subsection{Related Works and Motivation}
As the major objective in capacitive sensing is utilization, a variety of new applications have been designed and implemented upon prominent domains. Capacitive sensing can be categorized into two major fields: contact sensing and non-contact sensing. Diverse researches have been conducted in capacitive contact sensing [4, 5, 6, 7], and remarkable approaches have emerged that progressed the usability of capacitive sensing. The capacitive touch sensing technology has been successfully introduced as a low-cost and effective method to control the interface of IT devices and other controller systems. In addition, contact sensing has widely been utilized in the field of Health care [8] to monitor the electric signals from our body such as electrocardiogram [9, 10], electromyography [11], and electroencephalography. Moreover, some systems that detect a person's existence and compute the localization through capacitive sensing have been devised in [12-15]. Diverse studies were performed regarding hand motion detection, especially implementing visual processing techniques based on the footage or image recognition through image sensors [16-23]. Recently, as capacitive sensing tends to be more cost-effective in computational perspective compared to image sensing, the current research trends are shifting towards capacitive sensing to obtain an accurate and efficient methodology to recognize hand motions. Wong et. al. [6] designed a wearable capacitive sensing unit, sensing the capacitive values through attaching sensors to a hand and recognizing a specific hand gesture using a machine learning model. Similarly, Kalpattu et. al. [7] developed a gesture recognition glove device that utilizes capacitive sensors to extract voltage signals, perceiving a certain hand gesture. Likewise, most systems that detect hand motions imply recognizing static motions with certain hand forms such as thumbs-up pose, pouch pose, open-hand pose, or American sign language (ASL) [6, 7, 12, 24, 25]. \\
On the other hand, our hand gesture focuses on capturing dynamic motions in a non-contactive fashion such as waving a hand in different directions. Aezinia et. al. [26, 27] implemented three copper plates to measure the amplitude when a finger is nearby, and observed the EF difference around the plates, suggesting the feasibility of a motion tracking system through capacitive sensing. Arshad et. al. [28] conducted a study to track a person's existence through capacitive proximity sensing where the sensed signals contain distinct uniqueness which is sufficient enough to determine the existence of a specific person based on their experiments. Chu et. al. [29, 30] suggested a possibility of utilizing a neural network and Hidden Markov Model to recognize hand gesture EF signals. However, their study does not have thorough experimental validation, which lacks practical feasibility. Jung et. al. [31, 32] designed a signal offset computation method for accurate zero setting and calculated the empirical threshold of the EPS (Electric Potential Sensor) capacitive sensor. In addition, this study is based on our past studies [33, 34] which presented the dynamic threshold and frame extraction based on EPS capacitive sensor, classifying four-hand gesture types through the CNN model.

\subsection{System Overview}
Our system contains four capacitive sensors in the form of copper plates followed by middleware as shown in Fig.\ref{SystemOverview}. All the signal data and information from the sensors were collected using Arduino devices. Once the data are collected, they will be automatically passed through the pipeline of different pre-processing steps involving signal processing, signal detection, and frame extraction computations. Once the data processing pipeline finishes, pre-trained neural network models will determine the gesture type of the incoming authentic gesture dataset. As part of the final step for any kind of inference or usage of this system in the real-time scenario, we implemented the system into a python-based GUI framework that is being controlled based on the classification output of the neural network for an activity near the sensors.

\section{Problem Formulation and Methodology}
\subsection{Computing Optimal Threshold through Signal Analysis}
In this section, we analyze the extracted raw sensor signals and compute the optimal threshold by interpreting the internal properties. As EF signals can be easily influenced by the surrounding environment which causes unstable disturbances or noises in our data collection, our primary objective is to detect the relevant signals by reducing the intrinsic noises as humanly as possible. In general, there are two main perspectives to diminish noises: a) hardwarical approach such as designing the circuital structure b) applying the mathematical filtering computation that could magnify the target signals more distinctive based on software perspective. In the following sections, our aim is to discover the optimal softwarical approach to effectively avoid noises, suggesting an algorithmic resolution based on property analysis.

\subsection{Capacitive Theories}
In order to design a comprehensive framework by interpreting the fundamental background, comprehending the physical theories behind the capacitive sensing is essential [1]. A capacitive sensor measures the mutual capacitance interacting with nearby objects. Let $s=\left\{\left(\boldsymbol{s}_{i}\right)\right\}_{i=1}^{N}
$ be a set of sensors where $n(\bigcup_{\forall i}s_{i})=N$. The total self-capacitance $\mathrm{C}=\sum_{\forall i}\mathrm{c}_{i}$, where $\mathrm{c}_{i}$ indicates each sensor's (${s}_{i}$) capacitance since they are connected in parallel, and based on $Q_{(o,s)}=\mathrm{C}V$, $\mathrm{C}=\frac{\epsilon A_{s}}{d_{(o,s)}}$. $\epsilon$ denotes the dielectric constant, $o$ is an object, $d$ indicates a distance and $A_{s}$ is the overlapping area between sensor $s$ and a nearby object. Our final measurement $V$ can be approximated to Equation (1) where $Q$ denotes the amount of charges.

\begin{equation}
{ V \approx \bigcup_{\forall s} V_{s} \,\,\, s.t. \,\,\, V_{s} \approx \frac{Q_{(o,s)}}{ \frac{ \epsilon \cdot A_{s}}{d_{(o,s)}}}  }
\end{equation}

\subsection{Signal Processing}

\begin{enumerate}
\item  \textbf{Signal Analysis}\\
Let $x_i$ be our raw input signal. $X_s=\{x_{\left(s,i\right)}|1\le s\le 4,\ (s,i\mathrm{)}\in \mathbb{N}\mathrm{\}}$ and $I=\{i|1\le i\le n(X_s)\}$ where $n\left(X_{\exists s}\right)=n(I)$, and $i$ is an index of input $x$. Based on our empirical signal observation, the internal range of $\left|x_{(s,i)}\right|$ is $\left|x_{(s,i)}\right|-C\le \left|x_{\left(s,i\right)}\right|\le \left|x_{\left(s,i\right)}\right|+C$, with $C=\sigma \left(\left|x_{\left(s,i\right)}\right|\right)\approx 2.03$ when there are no disturbances from the surrounding environment, where $\sigma$ denotes the standard deviation and the unit is voltage $V$. Compared with this inherent oscillation, when a certain hand gesture takes place near a sensor with the vertical height of 0.1$\mathrm{\sim}$15cm, the signal amplitude has the form of non-linear relation between the distance and amplitude based on the coulomb's law $F=k_e\frac{|q_{1}q_{2}|}{r^{2}}$, where $k_e$ is the Coulomb's constant, $q_1$ and $q_2$ denote the magnitudes of charges, and $r$ is the distance between $q_{1}$ and $q_{2}$. Another property of a capacitive sensor is that it periodically discharges the internal charges as the sensor absorbs the electrons through nearby substances such as air and objects, when it surpasses the intrinsic capacity $\omega_s$, having $|x_s|=\sum_{i=1}^p|q_{i}|$, where $p$ indicates a certain period, and at $p+1$, $x_{(x,i=p+1)}=\sum_{i=1}^q|q_{i}-\omega_s|$. The relation of $x_s$ is as follows in Equation (2).

\begin{equation}
  \label{eqn:capactive}
|x_{(s,i^{'})}|\ll |x_{(s,i^{'}+p)}| \,\, and \,\, |x_{(s,i^{'}+p)}|\gg |x_{(s,i^{'}+p+1)}| \,\, s.t. \,\, |x_{(s,i+p)}| \approx \omega_s \,\, and \,\, |x_{(s,i+p)}|>\omega_s
\end{equation}

In Equation (2), $i+p$ indicates the moment of the discharge, and $i+p+1$ denotes the certain period after the discharge. $p$ is determined based on the variation of signal $\Delta x_s$, and our system displays $p\approx 1499.4\ \left(\pm 7.65\right),\ $that is $19.6\ (\pm 0.10)$ seconds when the sampling rate is 76.5 Hz, with $\sigma \left|x_{(s,i^{'})}\right|\approx \mathrm{11}\pm 4.73$. The reason for $p_{\forall s}$ are identical is that $A_{\forall s}$ are equivalent. However, $X_{\forall s}$ have their own values, with $\mu \left(X_{s}\right)\napprox \mu \left(X_{s^{'}}\right)$ where $s\neq s{'}$, but $\sigma \left(X_{s}\right)\approx \sigma \left(X_{s^{'}}\right)$, which implies that the overall range of difference is equivalent. Another property is that $x_{\left(s,i\right)}$ starts off from its distinct range, indicating the potential energy it possesses. Furthermore, $\mu \left({\tilde{X}}_{\left(s,j\neq j^{'}\right)}\right)\napprox \mu \left({\tilde{X}}_{\left(s,j^{'}\neq j\right)}\right)$ which entails that the overall amplitude of signal amplitude alters throughout the time where $\tilde{X}\mathrm{\in }X$, having $X=\bigcup_{\forall j}{{\tilde{X}}_j}$, and $j$ and $j^{'}$ are both indexes of $\tilde{X}$. Due to these diversities, setting the average values to zero is constantly required to reduce heterogeneity, by computing the offset ${\lambda}_s$ of each $s$, which we will introduce in Section 3.1.

\item  \textbf{Frequency Analysis}\\
By mapping our time-series dataset into the frequency domain, it offers the intuition of comprehending the intrinsic frequency components and insight to manipulate the frequency range using filtering mechanisms. Using the Fast Fourier Transform (FFT) in Equation (3), where \textit{f(x)} is an input signal, \textit{F(u)} is a linear aggregation of continuous periodical function ${cos2\pi ux + vsin2\pi ux}$, a term encompassing $v$ denotes an imaginary number, and \textit{u} indicates a frequency. Fig.\ref{FFT}(a) displays the FFT decomposition result when no hand gesture is presented and Fig.\ref{FFT}(b) shows the transformed signal when multiple hand gestures are presented.

\begin{equation}
{F(u) = \int^{\infty}_{-\infty} f(x) ({cos2\pi ux + vsin2\pi ux}) dx }
\end{equation}

\begin{figure}%
    \centering
    \subfloat[\centering ]{{\includegraphics[width=6.5cm]{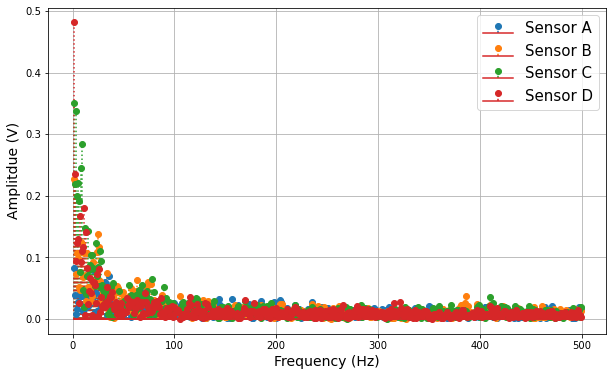} }}%
    \subfloat[\centering ]{{\includegraphics[width=6.5cm]{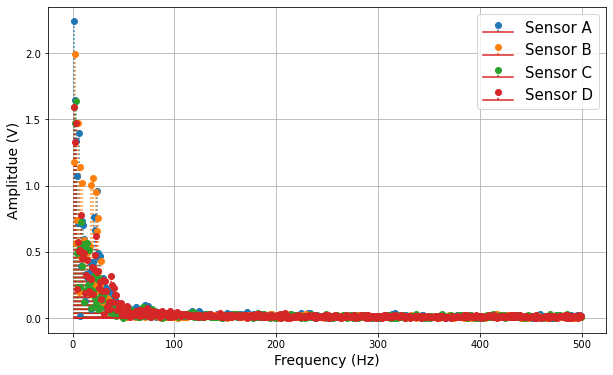} }}%
    \caption{FFT result: (a) without hand gestures; (b) with hand gestures (Left to Right).}
    \label{FFT}%
    
\end{figure}

The result in Fig.\ref{FFT} indicates that both raw signal without hand gesture (a), and the hand motion gesture signal (b) contain a significant amount of amplitudes in frequency range of under 50Hz. When computing the statistics: mean ($\mu (\cdot)$) and standard deviation ($\sigma (\cdot)$) by the different ranges of frequencies, where $\frac{1}{4} \sum_{s=1}^{4} (\sigma(x_{(s,j)}))$, $\frac{1}{4} \sum_{s=1}^{4}(\mu(x_{s,j}))$ in frequency range of $[a_n,b_n]$ are given in Table 1, with $\bigcup^{5}_{n=1}{a_{n}} = \{1,100,200,300,400|a_{1 \leq n \leq 5} \}$ and $\bigcup^{5}_{n=1}{b_{n}} = \{100,200,300,400,500|b_{1 \leq n \leq 5} \}$. For example, 1:100 in the first row in Table 1 indicates the frequency range of [1,100] Hz, and its statistics in the identical column below are the values after the band-pass filter where the passband is [1, 100] Hz.

\begin{table}
\caption{Statistical comparison}
\setlength{\tabcolsep}{4.5pt}
\begin{center}
\begin{tabular}{c  c  c c c c c }    
\hline \hline
Statistic & Gesture & 1:100 & 100:200 & 200:300 & 300:400 & 400:500 \\ [0.75ex] 
\hline
$\mu(\hat{x}_{(s,j)})$ & F & 7.3 ($\pm2.8$) & 1.1 ($\pm0.1$) & 1.1 ($\pm0.1$) & 0.8 ($\pm0.1$) & 0.8 ($\pm0.1$) \\ 
\hline
$\sigma(\hat{x}_{(s,j)})$ & F & 16.2 ($\pm4.7$) & 0.7 ($\pm0.1$) & 0.9 ($\pm0.1$) & 0.5 ($\pm0.1$) & 0.5 ($\pm0.1$)   \\ 
\hline
$\mu(\hat{x}_{(s,j)})$ & T & 16.59 ($\pm2.0$) & 1.4 ($\pm0.1$) & 1.3 ($\pm0.1$) & 0.9 ($\pm0.0$) & 0.9 ($\pm0.0$)   \\ 
\hline
$\sigma(\hat{x}_{(s,j)})$ & T & 35.95 ($\pm6.3$) & 0.7 ($\pm0.1$) & 1.0 ($\pm0.0$) & 0.5 ($\pm0.0$) & 0.5 ($\pm0.0$)   \\ [0.75ex] 
\hline \hline
\end{tabular}
\end{center}
\end{table}

\item \textbf{Processing Input Signal}\\
In order to effectively reduce internal noises, we consider the following three signal processing methods: Low Pass Filter (LPF), differentiation by sensors, and differentiation by sequential time step. As internal frequencies of input signals are usually dense in the range of ultra-low frequency, filtering the signals with LPF that has 50 cut-off frequency $f$ is a practical technique, which permits  $F(u\leq f)$ in Equation (1) and reverts by $\sum^{f}_{u=0} F(u) = f(\Bar{x}_{(s,j)})$ where $f(\Bar{x}_{(s,j)})\in f(x_{(s,j)})$. The major drawback of utilizing filters is that there is latency for the length of window size $w$, for $\frac{1}{w} \sum_{j=1}^{w} |x_{(s,j)}|=\hat{x}_{(s,j)}$, where $\hat{x}_{(s,j)} \in \hat{X}_s$ is our final processed signal. However, since the objective of our system is to control a UI, it requires a rapid response with low latency. Furthermore, this case requires an additional zero setting, which involves higher computation. The second option is to differentiate sensor signals in a pairwise manner, which is indicated in Equation (4) where $s \neq s^{'}$. This new $\hat{X}_{(s,s^{'})}$ comes effective when minimizing the noises, as sensors are proximal to each other and we assume that noises triggered by the external factors would affect almost identically among $\forall s$. We adjusted six combinations since $1\leq (s, s^{'}) \leq 4$, however, the results are not satisfactory as internal noises are still clearly shown.

\begin{equation}
{\hat{X}_{(s,s^{'})} = \bigcup_{\forall j} \hat{x}_{(s,s^{'},j)}  \,\,\, \textit{s.t.} \,\,\, \hat{x}_{(s,s^{'},j)} = |x_{(s,j)} - x_{(s^{'},j)}}|
\end{equation} 

The final option we inspect is shown in Equation (5), and our empirical study shows that this scheme is the most effective one among the three options. Equation (5) differentiates the sequential values between $x_j$ and $x_{(j-1)}$, and having $\hat{x}_{(s,j)} \approx 0$ when $\sigma^2(\bigcup_{j}^{j+p_{1}} x_{(s,j)})$ is low, providing automatic zero setting with low computations where $p_{1} \in \mathbb{N}$ is a constant that implies a certain period. Therefore, we select Equation (5) as our final processing mechanism, and add sensitivity $\tau_{s}$ and $1-\tau_{s}$ to each term $x_{(s,j)}$ and $x_{(s,j-1)}$ respectively to endow the corresponding weights, as well as moving the average to smooth out the values as shown in Equation (6) where $\Breve{x}_s$ indicates the processed signals, and $\Breve{X}_s$ denotes the set of $\forall \Breve{x}_{\exists s}$. Moreover, $\lambda_{s}$ indicates the offset of each $\exists s$, which is an auxiliary measure to set $\mu(\Breve{x}_{(s,j^{'} \leq j \leq j^{'}+p_{1})})\approx 0$, where $\lambda_{s} = \frac{1}{p_{1}} \sum_{j=j^{'}}^{j^{'}+p_{1}} \Breve{x}_{(s,j)}$. Fig.\ref{signal}(a) shows the raw signals $X_{\forall s}$ where Fig.\ref{signal}(b) shows the processed signals $\Breve{X}_{\forall s}$. 

\begin{equation}
{\hat{X}_{s} = \bigcup_{\forall j} \hat{x}_{(s,j)}  \,\,\, \textit{s.t.} \,\,\, \hat{x}_{(s,j)} = |x_{(s,j)} - x_{(s,j-1)}}|
\end{equation} 

\begin{equation}
\Breve{X}_s = \bigcup_{\forall j} (\Breve{x}_{(s,j)} - \lambda_{s}) \,\,\,\, \textit{s.t.} \,\,\, \Breve{x}_{(s,j)} = \frac{1}{p_{1}} \sum^{p_{1}}_{j} (\tau_{s} x_{(s,j)} + (1-\tau_{s}) x_{(s,j-1)})
\end{equation}

\begin{figure}%
    \centering
    \subfloat[\centering ]{{\includegraphics[width=6.5cm]{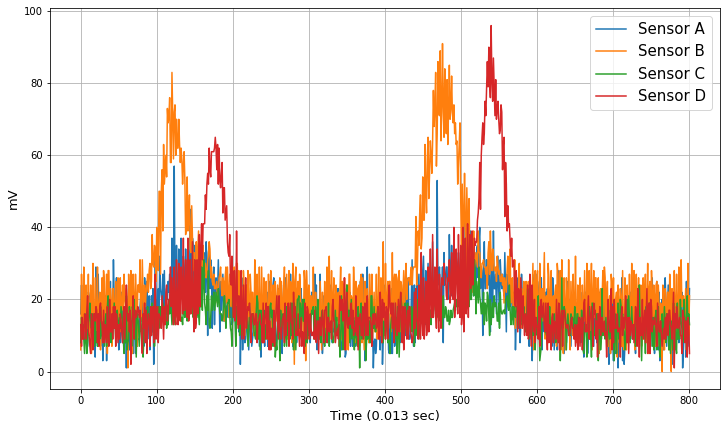} }}%
    \subfloat[\centering ]{{\includegraphics[width=6.5cm]{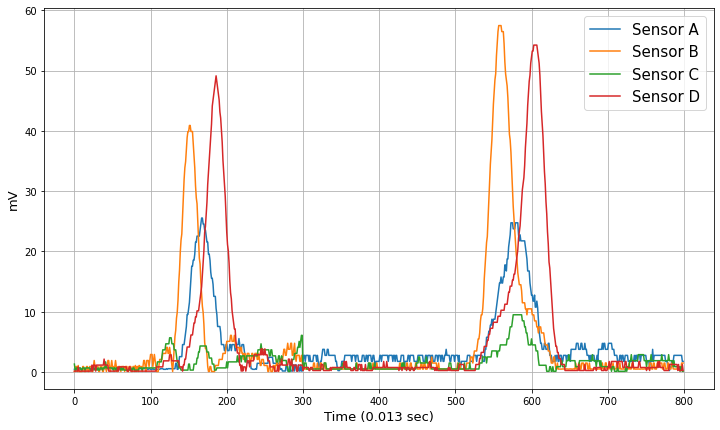} }}%
  \caption{(a) raw signals with a single hand gesture presented, (b) processed signals covering two hand gestures (Left to Right).}
    \label{signal}%
    
\end{figure}


\end{enumerate}

\subsection{Setting Adaptive Threshold}
As EF signals from sensors are highly variable throughout the extraction, setting an optimal threshold that could obtain accurate detection is essential. We suggest an adaptive threshold that changes dynamically, by periodically recomputing the threshold which can adapt to the inconsistent signal range and provide credible detection performance. The main purpose of the adaptive threshold is to detect the signal triggered by a hand gesture and to extract the frame that encompasses the authentic parts of the gesture. Let $\delta_{s}$ be a threshold of an individual sensor $s$ and $\phi$ be a constant that determines the amplitude of $\delta_{s}$. $\delta_{s}$ is being updated with period $p_{1}$ as shown in Equation (7). $\phi$ is decided with empirical observations, where $\phi$ directly influences the detection algorithm, determining the lower bound of $\delta_{s}$, such that $\inf \delta_{s} \approx \phi$ when base term $\frac{1}{p_{1}} \sum_{j=1}^{p_{1}} \Breve{x}_{(s,j)} \approx 0$. When $(\Breve{x}_{(s,j)} - \lambda_{s}) \geq \delta_{s}$ and $(\Breve{x}_{(s,j^{'} > j)} - \lambda_{s}) \leq \delta_{s}$ where $(j+50) \geq j^{'}$, the system recognizes this as a detection and extracts $\Breve{x}_{(s,j-p_{s}\leq j \leq j^{'}+p_{e})}$ where $p_{s}$ and $p_{e}$ are the constants that indicate period, in order to secure the whole frame that shifts forward from the start of the detection (i.e., $\Breve{x}_{(s,j)} - \lambda_{s} \geq \delta_{s}$) which is at index $j$, and moves backward from the end of the detection (i.e., $\Breve{x}_{(s,j^{'} > j)} - \lambda_{s} \leq \delta_{s}$), which is at index $j^{'}$. The system consists of two parts: offset initialization (Algorithm 1), and hand gesture detection through adaptive threshold that returns the frame automatically in real-time (Algorithm 2).

\begin{equation}
\delta_{s} = \frac{1}{p_{1}} \sum_{j=1}^{p_{1}} (\Breve{x}_{(s,j)} - \lambda_{s}) + \phi
\end{equation} 

\begin{algorithm}
\caption{Offset Initialization}\label{alg:cap}
\begin{algorithmic}\\
\textbf{Input: Initialize period $p_{0A}$} \\
\textbf{Output: Offset $\lambda_{s}$} \\

\For {each $s$} \textbf{in parallel} 
    \State Initialize $(start_{s}, end_{s}) \gets 0$
    \State Initialize $\delta_{s} \gets \phi$
    \State Initialize empty list $\Breve{X}_{s}, init_{s}$
    \For \, $j$ = 1, {\dots}, \, $p_{0A}$
        \State  $\Breve{X}_{s} \bigoplus \Breve{x}_{(s,j)}$ \Comment{append} 
    \EndFor
    \State $\lambda_{s} \gets \frac{1}{n(\Breve{X}_{s})} \cdot \sum_{\forall j} \Breve{x}_{(s,j)} $
    \State \textbf{Return} $\lambda_{s}, start_{s}, end_{s}, \delta_{s}, init_{s}$
\EndFor 
\end{algorithmic}
\end{algorithm}

Furthermore, in preparation of the case where sensor values suddenly surge when a dialectic object is being directly contacted by the sensor (huge amount of charges flow in this case compare with non-contact methods), additional measure to take precaution needs to be designed. When $|j-j^{'}| > p_{safe}$, the system perceives this outlier as malfunction and recomputes until $\Breve{X}_{s}$ is stabilized using offset $\lambda_{s}$. The overall steps of detection \& frame extraction are shown in Algorithm 2. Before running Algorithm 2, we compute the initial $\lambda_{s}$ for a certain period $p_{0}$ where $A \bigoplus B$ indicates the appending $B$ to $A$ function, and $start_{s}$ and $end_{s}$ denote the starting frame and ending frame respectively.

\begin{algorithm}[t]
\caption{Hand Gesture Detection through Adaptive Threshold}\label{alg:cap}
\begin{algorithmic}\\
\textbf{Input}: {input signal $\Breve{x}_{(s,j)}$, offset/threshold update period $p_{1}$, period that adds forward from the start $p_{s}$, period that adds backward from the end $p_{e}$, auxiliary period $p_{safe}$, auxiliary period $p_{0B}$ } \\
\textbf{Output: $FRAME$ } \\

\For {each $s$} \textbf{in parallel} 
    \State initialize empty list $frame_{s}$ 
\EndFor
\State $START, END \gets 0$
\State $(\lambda_{s}, start_{s}, end_{s}, \delta_{s}, init_{s}) \gets \text{Offset Initialization}$ \Comment{Algorithm 1}
\While {$\Breve{x}_{(s,j)} \neq \emptyset$} \Comment{run until input signal halts}
    \For {each $s$} \textbf{in parallel} 
        \State $\Breve{X}_{s} \bigoplus (\Breve{x}_{(s,j)}-\lambda_{s})$
        \If{$start_{s}$ = 0 and $end_{s}$ = 0} 
            \State $init_{s} \bigoplus \Breve{x}_{(s,j)}$  \Comment{Compute offset when frame is not set}
        \EndIf
        \If {$(\Breve{x}_{(x,j)} - \lambda_{s}) > \delta_{s}$ and $(\Breve{x}_{(s, j-1)} - \lambda_s) < \delta_{s}$}  \Comment{$\Breve{x}_{(s,j)}$ passed the $\delta_{s}$}
            \State $start_{s} \gets j - p_{s}$ \Comment{Shift forward to cover the authentic signal}
        \EndIf
        \State \textbf{else if} {$(\Breve{x}_{(s,j)}-\lambda_{s}) > \delta_s$} \textbf{then} \Comment{In the middle of higher than $\delta_{s}$} 
            \State $cnt \gets cnt + 1$  \Comment{Safety measure}
        \State \textbf{else if} {$(\Breve{x}_{(s,j)}-\lambda_{s}) < \delta_s$ and ($\Breve{x}_{(s,j-1)}-\lambda_{s}) > \delta_s$ } \textbf{then}
            \State $end_{s} \gets j + p_{e}$  \Comment{Shift backward to cover the authentic signal}
            \State $cnt \gets 0$
            
        \State \textbf{else if} {$start_{s} = 0$ and $end_{s} = 0$ and $j \% p_{1}=0$} \textbf{then}
            \State $\lambda_{s} \gets \mu(init_{s}) $ 
            \State $\delta_{s} \gets \frac{1}{p_{1}} \cdot \sum_{k=j-p_{1}}^{j} \Breve{x}_{(s,k)} + \phi$
            \State $init_s \gets empty list$
            
        \If{  $\Breve{x}_{(s,j)} > \delta_{s}$ and $cnt > p_{safe}$   } \Comment{safety measure}
            \State $\delta_{s} \gets \frac{1}{p_{1}} \cdot \sum_{k=j-p_{1}}^{j} \Breve{x}_{(s,k)}$ \Comment{offset recompute}
            \State $cnt \gets 0$
        \EndIf
        \If {$j = end_{s}$ and $j > p_{0B}$} \Comment{Another initial period}
            \State $frame_{s} \bigoplus array(start_{s}, end_{s})$
            \State $(start_{s}, end_{s}) \gets 0$
        \EndIf
    \EndFor
    
    \If {$p_{0B} < j$}
        \For {each $s$} \textbf{in parallel} 
            \If {$frame_{s} \neq \emptyset$}
                \State $START \gets frame_{s}[n(frame_{s})][0]$
                \State $END \gets frame_{s}[n(frame_{s})][1]$
            \EndIf
        \EndFor
        
        \If {$j = END$ and $START \neq 0$ and $END \neq 0$}
            \State $FRAME \bigoplus array(START, END)$
            \For {each $s$} \textbf{in parallel} 
                \State $frame_{s} \gets empty list$
                \EndFor
            \State $(START, END) \gets 0$
            \State return $FRAME$
        \EndIf
    \EndIf
\EndWhile

\end{algorithmic}
\end{algorithm}

In Algorithm 2, $init_{s}$ is a supplementary measure to compute $\lambda_{s}$ storing $\Breve{x}_{(s,j)}$ for $p_{1}$. In order to determine the final frame, let $START, END$ be a local volatile variable that denotes the temporary index of the start and end frame using $frame_{s}$. Then we append $array(START, END)$ to $FRAME$, having $FRAME[j][0] = START$ and $FRAME[j][1] = END$ where $j$ is the current index. We define the final frame as $\psi_{k}$, and its mathematical expression is shown in Equation (8) where $k$ is an index of the extracted frame. Note that $\lambda_{s}$ is iteratively being updated.

\begin{equation}
\psi_{k} = \bigcup_{j=Frame[k][0]}^{Frame[k][1]} (\Breve{x}_{(s,j)}-\lambda_{s})
\end{equation}

\begin{figure}%
    \centering
    \subfloat[\centering ]{{\includegraphics[height=4.5cm,width=6.5cm]{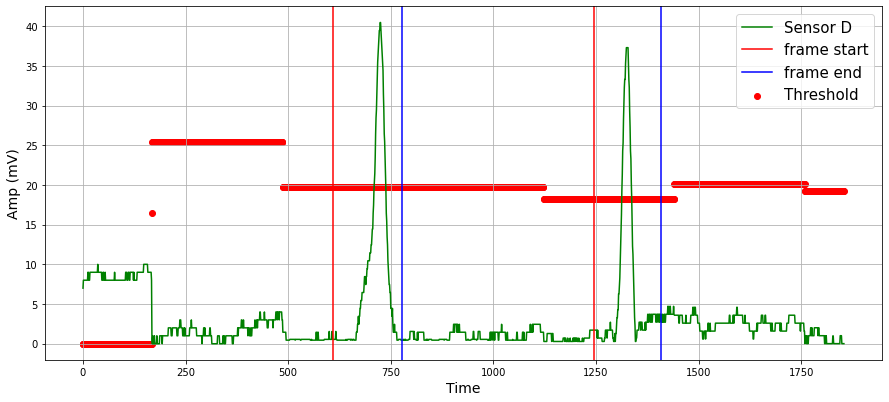} }}%
    \subfloat[\centering ]{{\includegraphics[height=4.5cm,width=6.5cm]{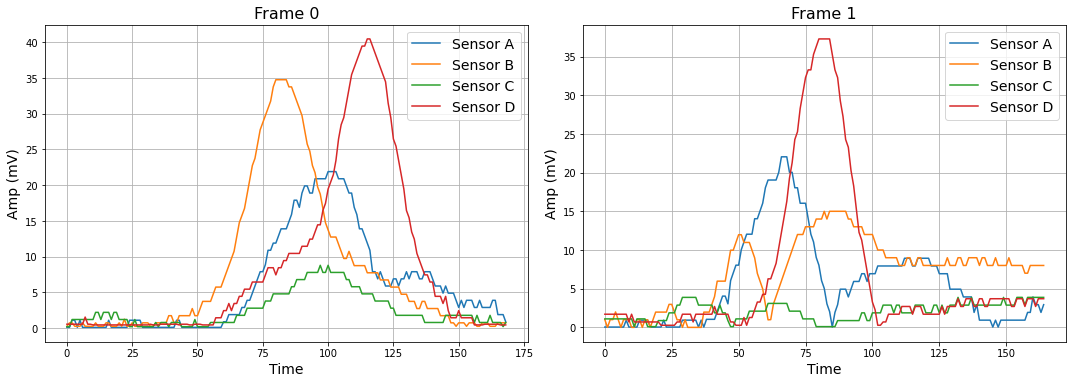} }}%
\caption{ (a) Adaptive threshold (red horizontal bars) which is periodically updated based on the amplitude of the input signal. (b) The frame extracted based on Algorithm 2.}
    \label{threshold}%
    
\end{figure}

Based on the designed algorithms, we automatically detect a hand gesture signal using an adaptive threshold and extract the signal frame that covers the authentic section triggered by the hand movement in real-time. The adaptive threshold and extracted motion frame are shown in Fig.\ref{threshold}. The parameter settings including multiple periods and the algorithm performances are demonstrated in Section 5.

\subsection{System Architecture and Components}
The hardware in our system includes three main components: four sensors (copper plates) for collecting EF signals, middleware (Arduino UNO) for aggregating sensor signals which transmits the signals to local devices through serial communication, and local devices. The sensors are located on top of the PVC plate which is made of non-conductive materials to preserve sensor signals. Recall that the input raw sensor signals are processed inside a device based on the methodologies in Section 3. Using the incoming signals, the system detects a motion gesture, extracts its frame, and classifies the frame implementing the algorithms proposed in Section 4. Finally, the classification result is transmitted through socket communication to a foreground application for UI control, which manipulates the interface with the built-in conditions such as {\it Left to Right} is to move to a next page and {\it Right to Left} is to go back to a previous page. The functions are listed in Table 2.

\section{Training Classifier and Controlling UI}
\subsection{Training AI Classifier}
As deep learning methods have been shown to have exceptional performance on predicting the input data into right class labels in a given feature space, we adopt a deep learning classifier model to train our hand gesture dataset to determine its class label among 10 gesture types. We employ the pre-trained classifier to conduct the classification of the extracted frames in real-time in order to utilize the prediction result as a conditional input signal to control UI. Our objective function in deep learning is shown in Equation (9) where $\mathscr{L}$ is a categorical cross entropy loss function with inputs of true label $y_{k}$, predicted label $g(\psi,M)$, and $M_{(s,k)}\ni {\{weight_{(s,k)}, bias_{(s,k)}\}}$ on frame index $k$. We update $M$ through gradient descent: $M^{(t+1)}_{(s,k)}=M^{(t)}_{(s,k)}-\eta \cdot \frac{\partial }{\partial M^{(t)}_{(s,k)}}\cdot {\mathscr{L}} (M^{(t)}_{(s,k)})$ where $\eta$ denotes the learning rate. 

\begin{equation}
{\mathop{\mathrm{min}}_{M_k\in {\mathbb{R}}^d}} \frac{1}{n(\psi_{(s,k)})}\sum_{\forall k}{{\mathscr{L}}(y_{(s,k)},g(\psi_{(s,k)},M_{(s,k})) } 
\end{equation} 

Since our dataset is in a time-series format, we implement two deep learning models which are proper for handling time-series data: Gated Recurrent Unit (GRU) and Long Short-Term Memory (LSTM). By training the neural networks, it returns a predicted label. The overall training performance is presented in Section 5-2.

\subsection{UI Control and Optimal System Structure}
The application UI is being directly controlled by a human in a foreground process, therefore the overall process including signal processing, thresholding, frame extraction, and classification should operate on the background process. The daemon process transmits the classification result to the foreground application through socket communication inside the single local device (Intel i7-1165G7, 2.8GHz, 16GB RAM). Depending on the gesture types, each has its certain function that controls a given UI. For example, when the classifier recognizes the hand gesture signal as ‘Left to Right’, then it commands to move to the next page. The other functions of motion class labels are shown in Table 2. We show a simple UI that visually shows through the window which contains the small GUI elements. Through importing the python GUI library Tkinter, we construct a window containing strings that signify the current hand motion gesture, for example, showing up “Left to Right” sending the input to the UI process. Again, this is to show the possibility of further implementation of diverse hand motion gesture class label inputs, which can be utilized to effectively control the target device.

\begin{table}
\caption{Types of hand motion gestures}
\setlength{\tabcolsep}{4.5pt}
\begin{center}
\begin{tabular}{c  c  c c c }    
\hline \hline
Left to Right & Right to Left & UP & DOWN & Down to Left \\ [0.75ex] 
\hline
Class 1 & Class 2 & Class 3 & Class 4 & Class 5 \\ 
Next page & Previous page & Scroll up & Scroll down & Previous 2 pages\\
\hline
Down to Right & Left to Down & Right to Down & Up to Left & Up to Right \\ 
\hline
Class 6 & Class 7 & Class 8 & Class 9 & Class 10 \\ Next 2 pages & Off & On & Volume down & Volume up\\
\hline 
\end{tabular}
\end{center}
\end{table}

\section{Experimental Evaluation}
\subsection{Experimental Setup}

The framework’s environmental setup consists of two main entities: four copper sensor plates for collecting EF signals, and middleware (Arduino UNO) for aggregating the raw sensor signal data which transmits the data to the local device through serial communication. Once the data have been collected, they will be passed through our pipeline as shown in Fig.\ref{SystemOverview} in which different pre-processing steps happen in a sequence. Once the data processing pipeline finishes, the processed signal is fed to the pre-trained neural network model for gesture classification. Finally, the classification result is transmitted through socket communication to another python GUI application process and the results of the gesture motions are shown on the interface for any activity that takes place near the sensors.

We conducted 10 different hand motion gestures which are listed in Table 2, and extracted the optimal signal frame. The following parameters are the designated parameters: $p_{s} = 70$, $p_{e} = 70$, $\phi = 20$, $p_{0A} = 10 ~seconds$, $p_{0B} = 8 ~seconds$, $p_{safe} = 3 ~seconds$, and $p_{1} = Sampling ~rate \cdot Initialization ~time$ where $Sampling ~rate = 53$ and $Initialization ~time = 6 ~seconds$. Based on the preset hyper-parameters, we operate our system and wave hand gesture motions near the sensor plates, measuring the classification accuracy of 10 gesture types and observing the accurate controllability of UI. Furthermore, we compare the accuracy and loss of two high-performance deep learning models: LSTM and GRU. For the layer structure of LSTM and GRU: LSTM (20) / GRU (20) – Dense (32) – Dense (64) – Dense (32) – Dense (10) with 60 epochs, batch size of 10, $\eta$ = 0.005, and relu activation function for the dense layers except for the final layer which uses the softmax.

\subsection{Experiment results}
\begin{figure}
    \centering
    \subfloat[\centering ]{{\includegraphics[width=6.5cm]{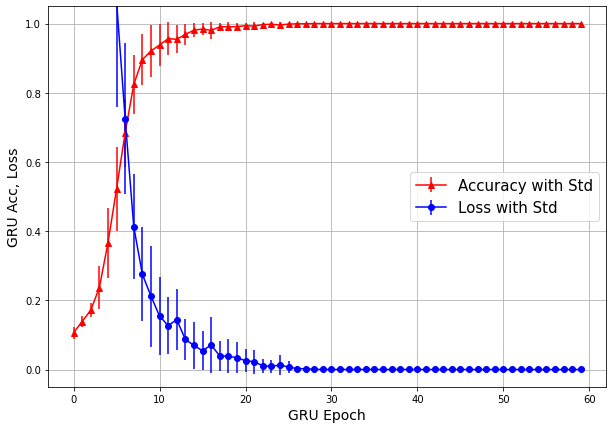} }}%
    \subfloat[\centering ]{{\includegraphics[width=6.5cm]{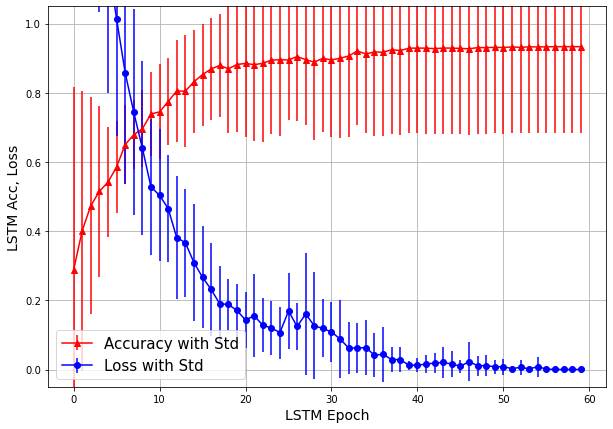} }}%
\caption{Performance of GRU and LSTM: (a) Average of GRU test accuracy and Loss with standard deviation after 30 training trials. (b) Average of LSTM test accuracy and Loss with standard deviation after 30 training trials.}
    \label{performance}%
    
\end{figure}

In order to evaluate our proposed framework, we use three evaluation metrics to assess our adaptive threshold, optimal period of the extracted frame, and train the ability of a deep learning classifier. The first metric is to quantitatively estimate the performance of the adaptive threshold in each sensor, which is solely for detecting the occurrence of hand gestures near the sensor plates and we define this as Detection rate. The second metric is to assess the coverage area of the extracted frame signal after detection, as the withdrawn frame is the key source for recognizing authentic hand motion types, and we denote this as Frame Extraction rate. Finally, in order to conduct accurate classification by calling the real-time hand gesture frames, a well-trained AI model is required, and the final metric is to evaluate the performance of the trainability of the classifier. Our overall evaluation shows 98.8\% Detection rate, 98.4\% Frame Extraction rate, and accuracy of 98.79 ($\pm0.72$)\% in GRU, 98.05 \%($\pm0.68$) in LSTM. Fig.\ref{performance} shows the average value of the validation dataset's accuracy and loss of pre-trained classifiers after 30 independent training trials, with standard deviation shown in each epoch step. The macro-average of precision, recall, and f1 score for all labels approximate to 1.0. In Fig.\ref{performance}, it shows GRU has more stable convergence with slightly higher accuracy at the end.
Therefore, we applied a pre-trained GRU classifier into the final background process and based on the classification result, it dynamically controls the foreground application process UI with 10 options. To quantitatively evaluate the overall performance of the suggested system, it operates with an average of 98.67 ($\pm 2.005$) \% accuracy.

\section{Conclusion and Future Work}
This paper proposes a real-time user interface control system based on hand motion gesture recognition through non-contact capacitive sensing. Our designed framework processes raw capacitance signal data from sensors with sequential differentiation, which are selected as an optimum option among filtering schemes with spectrum analysis, and two differentiation computations are performed after observing each result. By dynamically computing the adaptive threshold, the system detects a movement near sensors and automatically extracts the motion gesture frame that covers the authentic EF disturbance signal triggered by hand. This frame is sent to a pre-trained GRU classifier, which categorizes the input frame among 10 predefined hand motion gesture types. Furthermore, the classification result is transmitted to another application process that maneuvers the UI and controls the interface with built-in commands.  
This proposed work provides the feasibility of adopting a capacitive sensing system, which shows the possibility of commercializing intensive application products through capacitive proximity sensing as capacitive proximity sensing is emerging as a promising sensing technology that enables the implementation of new applications in the domains of IoT, VR, AR, touchless interface, HMI, robotics, HCI, and NUI, with cost-efficiency. The dataset used to support the study is uploaded in GitHub [3] and can be accessed by the public.

The limitations of our proposed system include limited distance between a sensor and a dialectic object, and unstable signal trend affected by the surrounding. Limited distance is an intrinsic drawback of a capacitive sensor as it is non-contact proximity sensing. The range of height for our system in order to detect a hand motion is approximately under 15cm. This distance may not be sufficient enough to be commercialized in products, therefore, more accurate and sensitive sensing is required. The overall trend of capacitive sensor signal differs when surrounding changes, and as we cannot predict the heterogeneous circumstances, this diversity is inevitable. This problem can partially gain a solution through two basic approaches, collecting diverse datasets and computing the physical model of the given environment in order to gain possible cases of how the signal pattern would occur. Our future work lies on the design of approaches to resolve the above mentioned limitations: analyzing the EF disturbance pattern when dialectic objects are near, and collecting more data. Moreover, we will focus on increasing the utility of the proposed system, adding further hand gesture motion data with an increasing number of hand gesture types (at least 100 gestures). After successfully classifying the labels and connecting more sensors to construct a more comprehensive system, these steps will contribute to constructing an intensive system with higher multiplicity that utilizes capacitive sensing.\\\\

\end{document}